\documentstyle[12pt]{article}

\newcommand {\e} {\mbox{\rm e}}

\newcommand {\nn}    {\nonumber}
\newcommand {\vs}[1]  { \vspace*{#1 cm} }

\newcounter{eq}
\newcounter{sc}

\newcommand {\PL}   {Phys.Lett.}
\newcommand {\PR}   {Phys.Rev.}
\newcommand {\PRL}   {Phys.Rev.Lett.}

\def\overleftrightarrow#1{\vbox{\ialign{##\crcr
 $\leftrightarrow$\crcr\noalign{\kern-1pt\nointerlineskip}
 $\hfil\displaystyle{#1}\hfil$\crcr}}}

\newlength{\minitwocolumn}
\begin{document}

\begin{flushright}
EDO-EP-37\\
March, 2001\\
\end{flushright}
\vspace{30pt}

\pagestyle{empty}
\baselineskip15pt

\begin{center}
{\large\bf Antisymmetric Tensor Fields in the Locally 
Localized Gravity Models

 \vskip 1mm
}

\vspace{20mm}

Ichiro Oda
          \footnote{
          E-mail address:\ ioda@edogawa-u.ac.jp
                  }
\\
\vspace{10mm}
          Edogawa University,
          474 Komaki, Nagareyama City, Chiba 270-0198, JAPAN \\

\end{center}


\vspace{15mm}
\begin{abstract}
We study the localization property of antisymmetric tensor fields
in the locally localized gravity models. It is shown that all
the antisymmetric tensor fields, including the vector field, in a 
bulk space-time are trapped on an $AdS$ brane by a gravitational 
interaction where the presence of the brane cosmological constant 
plays an important role as in the cases of the other bulk fields.
The normalized zero-modes spread rather widely in extra space so small
extra dimensions might be needed in order not to conflict with experiment.

\vspace{15mm}

\end{abstract}

\newpage
\pagestyle{plain}
\pagenumbering{arabic}

\rm

There has been a great deal of excitement recently over an alternative
compactification scenario of space-time, where our four dimensional
world emerges as a topological defect, a 3-brane, in a higher dimensional 
space-time \cite{Randall1, Randall2}. 
(This model was generalized to the case of many branes in 
Ref. \cite{Oda1}.)
In such 'brane world' models, it is sometimes supposed that all the
familiar matter and gauge fields are constrained to live on the brane,
whereas gravity is free to propagate in the whole space-time.
However, this treatment is not democratic at all as long as we
do not have any reasonable mechanism of confinement of the matter
and gauge fields on the brane. Thus, we should take account of all
the local fields as bulk fields in the 'brane world' models.

In particular, one crux in the original Randall-Sundrum model is then 
how to localize the bulk gauge fields on a brane by a gravitational 
interaction.
It is well known that in the Randall-Sundrum model the photon and the
Kalb-Ramond field cannot be localized on the brane \cite{Bajc, Oda3, Oda4}.

Recently, in Refs. \cite{Oda5, Oda6} we have developed a new 
mechanism for the localization of the bulk fields on the brane in the locally 
localized gravity models \cite{Kogan, Karch, Miemiec, Schwartz, Tachibana}. 
In this mechanism, the presence of the brane cosmological constant
plays a crucial role.  

The aim of this paper is to study the localization property of the bulk
antisymmetric tensor fields in the locally localized gravity models in detail.
In previous works \cite{Oda5, Oda6}, the antisymmetric tensor fields were not 
considered explicitly. But if we regard the models at hand as coming from
superstring theory a bunch of antisymmetric tensor fields are expected to
emerge in the low energy field theory so it is of importance to give a full 
analysis for the bulk antisymmetric tensor fields. Indeed, D-branes, in which
D3 brane is one of candidates as our world, carry the
nontrivial charges of the R-R antisymmetric tensor fields \cite{Pol}. 
Moreover, it is pointed out that a brane with the charge of an antisymmetric
tensor field might be stable owing to the charge conservation law
\cite{Gherghetta}. 

First, we shall start with a locally localized gravity model by Karch and
Randall in five dimensions \cite{Karch}. Unlike their paper, we shall make
use of the formulation adopted in our previous paper \cite{Oda6} since not
only
the translation to the Karch-Randall formulation but also the
generalization to 
higher dimensions are straightforward. 
The metric ansatz we take is of the form \cite{Oda6}:
\begin{eqnarray}
ds_{(5)}^2 &=& g_{MN} dx^M dx^N  \nn\\
&=& \e^{-A(r)} \hat{g}_{\mu\nu} dx^\mu dx^\nu + dr^2  \nn\\
&=& \e^{-A(z)} (\hat{g}_{\mu\nu} dx^\mu dx^\nu + dz^2), 
\label{1}
\end{eqnarray}
where $M, N, ...$ denote five-dimensional space-time indices and 
$\mu, \nu$, ...four-dimensional brane ones. The metric over the brane
$\hat{g}_{\mu\nu}$ denotes the four-dimensional anti-de Sitter metric.
Moreover, $e^{-A(r)}$, $e^{-A(z)}$ and the relation between the 'radial' 
$r$-coordinates and the 'conformal' $z$-coordinates are given by
\begin{eqnarray}
\e^{-A(r)} &=& \cosh^2 \omega r, \nn\\
\e^{-A(z)} &=& \frac{1}{\sin^2 \omega z}, \nn\\
\e^{\omega r} &=& \tan \frac{1}{2} \omega z,
\label{2}
\end{eqnarray}
where $\omega$ denotes $\sqrt{\frac{-\Lambda}{6}}$ with $\Lambda$
being the negative bulk cosmological constant. The brane cosmological
constant $\Lambda_{AdS}$ is then expressed in terms of the bulk 
cosmological constant $\Lambda$ via $\Lambda_{AdS} = \frac{1}{2} \Lambda$. 
In other words, this model describes an $AdS_4$ brane sitting at the 
origin $r=0$ in $AdS_5$. 
Notice that since the 'radial' coordinate $r$ runs from $0$ to $\infty$, 
this relation yields the range of $z$, which is $\frac{\pi}{2 \omega} 
\le z \le \frac{\pi}{\omega}$. 

In what follows, we shall consider from a 0-form potential, i.e., a
massless real scalar field, to a 3-form potential in five dimensions.
A 4-form potential is non-dynamical field, for which we shall not take
into consideration. 
The action of a 0-form potential $\Phi$ is given by
\begin{eqnarray}
S_0 = - \frac{1}{2} \int d^5 x \sqrt{-g} g^{M_1 N_1} F_{M_1} F_{N_1},
\label{3}
\end{eqnarray}
where $F_M = \partial_M \Phi$. The equation of motion becomes
\begin{eqnarray}
\partial_{M_1} (\sqrt{-g} g^{M_1 N_1} F_{N_1}) = 0.
\label{4}
\end{eqnarray}
Let us look for a solution with the form of 
\begin{eqnarray}
\Phi(x^M) = \phi(x^\mu) u(z),
\label{5}
\end{eqnarray}
where we assume the equation of motion for the brane field 
$\hat{\nabla}^\mu f_\mu = 0$ with being $f_\mu = \partial_\mu \phi$.
With the ansatz (\ref{5}), Eq. (\ref{4}) reduces to a single differential
equation for $u(z)$:
\begin{eqnarray}
\partial_z \left( \e^{- \frac{3}{2} A(z)} \partial_z u \right) = 0.
\label{6}
\end{eqnarray}
The general solution is easily found to be
\begin{eqnarray}
u(z) = - \frac{\alpha}{3 \omega} \cos \omega z (\sin^2 \omega z + 2)
+ \beta,
\label{7}
\end{eqnarray}
where $\alpha, \beta$ are integration constants.

Plugging the form of a solution (\ref{5}) into the starting action,
the action can be cast to
\begin{eqnarray}
S_0^{(0)} &=& - \frac{1}{2} \int d^5 x \sqrt{-g} g^{M_1 N_1} F^{(0)}_{M_1} 
F^{(0)}_{N_1} \nn\\
&=& - \frac{1}{2} \int d^4 x \sqrt{-\hat{g}} \int dz \e^{- \frac{3}{2} A(z)}
\left[ u^2(z) \hat{g}^{\mu\nu} \partial_\mu \phi \partial_\nu \phi 
+ (\partial_z u)^2 \phi^2 \right].
\label{8}
\end{eqnarray}
Provided that the general solution (\ref{7}) is inserted to this action,
it turns out that the integral
\begin{eqnarray}
I_1 = \int_{\frac{\pi}{2 \omega}}^{\frac{\pi}{\omega}} dz 
\e^{- \frac{3}{2} A(z)} u^2(z) 
\label{9}
\end{eqnarray}
is in general divergent, but only when $\beta = - \frac{2 \alpha}{3 \omega}$
it becomes strictly finite. Then, the zero-mode $u(z)$ and $I_1$ are
respectively given by
\begin{eqnarray}
u(z) &=& - \frac{\alpha}{3 \omega} \left[\cos \omega z (\sin^2 \omega z
+ 2) + 2 \right], \nn\\
I_1 &=& \frac{\alpha^2}{9 \omega^3}(\frac{1}{3} + 4 \log 2). 
\label{10}
\end{eqnarray}
The fact that $I_1$ is finite shows that a 0-form potential in a bulk
is trapped on an $AdS_4$ brane by a gravitational interaction.

After evaluating the second integral over $z$ in (\ref{8}), we
obtain
\begin{eqnarray}
S_0^{(0)} = - \frac{1}{2} \int d^4 x \sqrt{-\hat{g}} 
\left[ \frac{\alpha^2}{9 \omega^3}(\frac{1}{3} + 4 \log 2)
\hat{g}^{\mu\nu} \partial_\mu \phi \partial_\nu \phi 
+ \frac{2 \alpha^2}{3 \omega} \phi^2 \right].
\label{11}
\end{eqnarray}
To make the kinetic term take a canonical form, let us redefine the field
as $\frac{\alpha}{3 \omega^{\frac{3}{2}}} \sqrt{\frac{1}{3} + 4 \log 2}
\phi \rightarrow \phi$. Then, we have
\begin{eqnarray}
S_0^{(0)} = - \frac{1}{2} \int d^4 x \sqrt{-\hat{g}} 
\left[ \hat{g}^{\mu\nu} \partial_\mu \phi \partial_\nu \phi 
+ \frac{6 \omega^2}{\frac{1}{3} + 4 \log 2} \phi^2 \right].
\label{12}
\end{eqnarray}
Since the brane cosmological constant $\Lambda_{AdS}$ is required to 
be tiny in order to make contact with observations, the mass of a scalar 
field on a brane must be very small.

At this stage, let us examine the zero mode $u(z)$ in more detail. 
First, note that $u(z)$ in (\ref{10}) satisfies the Dirichlet boundary 
condition at $z = \frac{\pi}{\omega}$ where it is 
known that the gravitational potential becomes infinity \cite{Karch}.
Second, we see that the normalized wave function in the $\hat{g}_{MN}$ 
space takes the form
\begin{eqnarray}
\hat{u}(z) &=& - \frac{1}{\sqrt{I_1}} \e^{-\frac{3}{4}A(z)} u(z) \nn\\
&=& \sqrt{\frac{\omega}{\frac{1}{3} + 4 \log 2}} \frac{-\cos^3 \omega z
+ 3 \cos \omega z + 2}{(\sin \omega z)^{\frac{3}{2}}} \nn\\
&=& \sqrt{\frac{2 \omega}{\frac{1}{3} + 4 \log 2}} 
\frac{3 \e^{-\omega r} + \e^{ -3 \omega r}}{(\e^{\omega r} + 
\e^{-\omega r})^{\frac{3}{2}}} \nn\\
& \approx& 3 \sqrt{\frac{2 \omega}{\frac{1}{3} + 4 \log 2}}
\e^{-\frac{5}{2} \omega r},
\label{13}
\end{eqnarray}
where  Eq. (\ref{2}) was utilized and the last expression holds
for $r \gg 1$. Eq. (\ref{13}) implies that the wave function is
localized near the origin $r = 0$. However, there is a caveat.
Because of $\omega \ll 1$ this wave function spreads rather widely
in a bulk space, so we might require the extra space to be small
enough to be consistent with observations. This 'small extra dimensions' 
scenario might shed light on the conventional Kaluza-Klein compactification 
idea. Namely, in the model at hand the size of extra dimension is
dependent on the size of the brane cosmological constant. As is well
known that observations suggest a very tiny cosmological constant,
extra dimension should be very small. At present it is not clear
whether this new scenario is really a viable one, but it is of interest
to note that the size of extra space is determined by that of
the cosmological constant on our brane.  
As a final remark, it is known that traceless, transverse graviton
modes in general obey the same equations of motion as a massless
scalar field in a curved background. Therefore, the present result
tells us that the graviton is also trapped on an $AdS_4$ brane.

Next, we shall turn to a 1-form potential, that is, the massless
$U(1)$ gauge field \cite{Oda5}. The path of arguments is very
similar to the case of a 0-form potential. The action is 
\begin{eqnarray}
S_1 = -\frac{1}{4} \int d^5 x \sqrt{-g} g^{M_1 N_1} g^{M_2 N_2} 
F_{M_1 M_2} F_{N_1 N_2}, 
\label{14}
\end{eqnarray}
where $F_{MN} = \partial_M A_N - \partial_N A_M$. The equations of motion
become 
\begin{eqnarray}
\partial_{M_1} (\sqrt{-g} g^{M_1 N_1} g^{M_2 N_2} F_{N_1 N_2}) = 0. 
\label{15}
\end{eqnarray}
After taking the gauge condition $A_z(x^M) = 0$, we search for a
solution with the form of
\begin{eqnarray}
A_\mu(x^M) = a_\mu(x^\lambda) u(z),
\label{16}
\end{eqnarray}
where we assume the equations of motion on a brane
$\hat{\nabla}^\mu a_\mu = \hat{\nabla}^\mu f_{\mu\nu} = 0$ with the
definition of $f_{\mu\nu} = \partial_\mu a_\nu - \partial_\nu a_\mu$.
With this ansatz, Eq. (\ref{15}) also reduces to a single differential
equation for $u(z)$:
\begin{eqnarray}
\partial_z \left( \e^{- \frac{1}{2} A(z)} \partial_z u \right) = 0,
\label{17}
\end{eqnarray}
whose general solution is given by
\begin{eqnarray}
u(z) = - \frac{\alpha}{\omega} \cos \omega z + \beta,
\label{18}
\end{eqnarray}
where $\alpha, \beta$ are integration constants.

Plugging this solution into the classical action (\ref{14}),
selecting $\beta = - \frac{\alpha}{\omega}$ which gives rise to only the 
finite kinetic term, and evaluating the integrals over $z$, the action 
takes the form
\begin{eqnarray}
S_1^{(0)} &=& -\frac{1}{4} \int d^5 x \sqrt{-g} g^{M_1 N_1} g^{M_2 N_2} 
F^{(0)}_{M_1 M_2} F^{(0)}_{N_1 N_2} \nn\\
&=& - \frac{1}{4} \int d^4 x \sqrt{-\hat{g}} 
\left[ \frac{\alpha^2}{\omega^3}(-1 + 2 \log 2)
\hat{g}^{\mu_1 \nu_1} \hat{g}^{\mu_2 \nu_2} f_{\mu_1 \mu_2} f_{\nu_1 \nu_2} 
+ \frac{2 \alpha^2}{\omega} \hat{g}^{\mu_1 \nu_1} a_{\mu_1} a_{\nu_1}
\right]. 
\label{19}
\end{eqnarray}
Redefining $\frac{\alpha}{\omega^{\frac{3}{2}}} \sqrt{-1 + 2 \log 2} a_\mu
\rightarrow a_\mu$, the above action can be rewritten as
\begin{eqnarray}
S_1^{(0)} = - \frac{1}{4} \int d^4 x \sqrt{-\hat{g}} 
\left[ \hat{g}^{\mu_1 \nu_1} \hat{g}^{\mu_2 \nu_2} f_{\mu_1 \mu_2} f_{\nu_1
\nu_2} 
+ \frac{2 \omega^2}{-1 + 2 \log 2} \hat{g}^{\mu_1 \nu_1} a_{\mu_1}
a_{\nu_1} \right]. 
\label{20}
\end{eqnarray}
As stressed in Ref. \cite{Oda5}, the massless condition of '$\it{photon}$'
on a brane forces us to choose $\omega \approx 0$ as expected from
a small value of the cosmological constant.

Let us study the zero-mode $u(z)$. With a specific choice $\beta 
= - \frac{\alpha}{\omega}$, $u(z)$ becomes
\begin{eqnarray}
u(z) = - \frac{2 \alpha}{\omega} \cos^2 \frac{1}{2} \omega z,
\label{21}
\end{eqnarray}
which also satisfies the Dirichlet boundary condition at $z =
\frac{\pi}{\omega}$ 
\cite{Oda5, Oda6}. And the normalized wave function is calculated
as
\begin{eqnarray}
\hat{u}(z) &=& - \frac{1}{\sqrt{I_1}} \e^{-\frac{1}{4}A(z)} u(z) \nn\\
&=& \sqrt{\frac{\omega}{-1 + 2 \log 2}} \frac{\cos \omega z + 1}
{(\sin \omega z)^{\frac{1}{2}}} \nn\\
&=& \sqrt{\frac{2 \omega}{-1 + 2 \log 2}} 
\frac{\e^{-\omega r}}{(\e^{\omega r} + \e^{-\omega r})^{\frac{1}{2}}} \nn\\
& \approx& \sqrt{\frac{2 \omega}{-1 + 2 \log 2}}
\e^{-\frac{3}{2} \omega r},
\label{22}
\end{eqnarray}
where  this time $I_1 =  \frac{\alpha^2}{\omega^3}(-1 + 2 \log 2)$ 
and the last expression also holds for $r \gg 1$. Eq. (\ref{22}) again
means that the wave function is localized near the origin $r = 0$
but spreads rather widely in a bulk space. It is remarkable that
both the 0-form potential and the 1-form potential share the
same features and demand the 'small extra dimensions' scenario.
Incidentally, $\hat{u}(z)$ in (\ref{13}) and (\ref{22}) also satisfies
the Dirichlet boundary condition at $z = \frac{\pi}{\omega}$. 

We are now ready to consider a 2-form potential, in other words, the
Kalb-Ramond field. In five dimensions, the 2-form potential is
dual to the 1-form potential, so it is expected that the
2-form potential is also trapped on a brane as in the 1-form one.
Below we shall explicitly show that this is indeed the case.
The arguments proceed in the same way as in the 0-form and 1-form potentials.
The classical action is given by 
\begin{eqnarray}
S_2 = -\frac{1}{12} \int d^5 x \sqrt{-g} g^{M_1 N_1} g^{M_2 N_2} 
g^{M_3 N_3} F_{M_1 M_2 M_3} F_{N_1 N_2 N_3}, 
\label{23}
\end{eqnarray}
where $F_{MNP} = 3 \partial_{[M} A_{NP]} = \partial_M A_{NP} + 
\partial_N A_{PM} + \partial_P A_{MN}$. The equations of motion
are 
\begin{eqnarray}
\partial_{M_1} (\sqrt{-g} g^{M_1 N_1} g^{M_2 N_2} g^{M_3 N_3}
F_{N_1 N_2 N_3}) = 0. 
\label{24}
\end{eqnarray}
The action (\ref{23}) has the following gauge symmetries and 
the first-stage off-shell reducible symmetry \cite{Batalin}
\begin{eqnarray}
\delta A_{MN} &=& \partial_M \varepsilon_N - \partial_N \varepsilon_M, \nn\\
\delta \varepsilon_M &=& \partial_M \varepsilon,
\label{25}
\end{eqnarray}
so we can take the gauge conditions $A_{Mz}=0$. (Note that the number
of degrees of freedom associated with the symmetries (\ref{25}) is
$5 - 1 =4$, which exactly coincides with the number of gauge conditions
$A_{Mz}=0$.) 
The natural ansatz for a solution is 
\begin{eqnarray}
A_{\mu\nu}(x^M) = a_{\mu\nu}(x^\lambda) u(z),
\label{26}
\end{eqnarray}
where the equations on a brane $\hat{\nabla}^\mu a_{\mu\nu} 
= \hat{\nabla}^\mu f_{\mu\nu\rho} = 0$ with the
definition of $f_{\mu\nu\rho} = 3 \partial_{[\mu} a_{\nu\rho]}$
are imposed.
With this ansatz, Eq. (\ref{24}) also reduces to a single differential
equation for $u(z)$:
\begin{eqnarray}
\partial_z \left( \e^{ \frac{1}{2} A(z)} \partial_z u \right) = 0.
\label{27}
\end{eqnarray}

At this stage, notice the similarity of this equation to the other
cases treated thus far. Given a p-form potential, the corresponding 
equation generally takes the form
\begin{eqnarray}
\partial_z \left( \e^{ - \frac{3-2p}{2} A(z)} \partial_z u \right) = 0.
\label{28}
\end{eqnarray}
This regularity stems from the number of $g^{MN}$ in the equations 
of motion, or equivalently in the action. In the 2-form potential,
compared to the lower form potentials, one notable change has
happened. Namely, the sign of the coefficient in front of $A(z)$
in the exponential has become positive in the present case, while
negative in 0-form and 1-form potentials. Due to this change,
it turns out that we cannot find a normalizable zero-mode $u(z)$ 
with the nontrivial dependence of a fifth dimension $z$. The 
only normalizable zero-mode is given by the constant mode: 
\begin{eqnarray}
u(z) = u_0 = const.
\label{29}
\end{eqnarray}
Substituting this solution into the action (\ref{23}), the action 
reduces to the form
\begin{eqnarray}
S_2^{(0)} = - \frac{1}{12} \frac{u_0^2}{\omega} 
\int d^4 x \sqrt{-\hat{g}} 
\hat{g}^{\mu_1 \nu_1} \hat{g}^{\mu_2 \nu_2} \hat{g}^{\mu_3 \nu_3}
f_{\mu_1 \mu_2 \mu_3} f_{\nu_1 \nu_2 \nu_3} 
\label{30}
\end{eqnarray}
Note that the constancy of the zero-mode leads to the vanishing mass 
term.

It is natural to ask ourselves if there is a critical difference between
a p-form and its dual (3-p)-form. For instance, at first sight, the form of 
the zero-mode $u(z)$ appears to be different between the two forms. 
But this is an illusion. Actually, it is the normalized zero-mode that
we should pay attention. In the case at hand, it is given by
\begin{eqnarray}
\hat{u}(z) &=& \frac{1}{\sqrt{I_1}} \e^{\frac{1}{4}A(z)} u(z) \nn\\
&=& \sqrt{\omega} (\sin \omega z)^{\frac{1}{2}} \nn\\
&=& \sqrt{2 \omega} \frac{1}{(\e^{\omega r} + \e^{-\omega r})^{\frac{1}{2}}}
\nn\\
& \approx& \sqrt{2 \omega} \e^{- {\frac{1}{2}} \omega r},
\label{31}
\end{eqnarray}
where $I_1$ is defined as $\frac{u_0^2}{\omega}$. As before, this
normalized wave function indeed satisfies the Dirichlet boundary condition at 
$z = \frac{\pi}{\omega}$ and has an exponentially decreasing form for
$r \gg 1$. 

The remaining form is a 3-form potential, for which we shall now study. 
The result is very similar to the case of a 2-form potential, so
we shall touch on this case.
The classical action of the 3-form potential is  
\begin{eqnarray}
S_3 = -\frac{1}{48} \int d^5 x \sqrt{-g} g^{M_1 N_1} g^{M_2 N_2} 
g^{M_3 N_3} g^{M_4 N_4} F_{M_1 M_2 M_3 M_4} F_{N_1 N_2 N_3 N_4}, 
\label{32}
\end{eqnarray}
where $F_{MNPQ} = 4 \partial_{[M} A_{NPQ]} = \partial_M A_{NPQ} - 
\partial_N A_{MPQ} + \partial_P A_{MNQ} - \partial_Q A_{MNP}$. 
The equations of motion are then 
\begin{eqnarray}
\partial_{M_1} (\sqrt{-g} g^{M_1 N_1} g^{M_2 N_2} g^{M_3 N_3} g^{M_4 N_4}
F_{N_1 N_2 N_3 N_4}) = 0. 
\label{33}
\end{eqnarray}
Taking the gauge conditions $A_{MNz}=0$ \footnote{These gauge conditions
fix gauge symmetries, 1st-stage, and 2nd-stage reducible symmetries
\cite{Batalin}, which are given by 
$\delta A_{MNP} = \partial_{[M} \varepsilon_{NP]},
\delta \varepsilon_{MN} = \partial_{[M} \varepsilon_{N]},
\delta \varepsilon_M = \partial_M \varepsilon$.} , and the ansatz  
\begin{eqnarray}
A_{\mu\nu\rho}(x^M) = a_{\mu\nu\rho}(x^\lambda) u(z),
\label{34}
\end{eqnarray}
with the equations on a brane $\hat{\nabla}^\mu a_{\mu\nu\rho} 
= \hat{\nabla}^\mu f_{\mu\nu\rho\sigma} = 0$ where 
$f_{\mu\nu\rho\sigma} = 4 \partial_{[\mu} a_{\nu\rho\sigma]}$,
Eq. (\ref{33}) becomes
\begin{eqnarray}
\partial_z \left( \e^{ \frac{3}{2} A(z)} \partial_z u \right) = 0.
\label{35}
\end{eqnarray}
As a normalizable solution, we also have to select the solution (\ref{29}),
{}from which the classical action can be written as
\begin{eqnarray}
S_3^{(0)} = - \frac{1}{48} \frac {2}{3} \frac{u_0^2}{\omega} 
\int d^4 x \sqrt{-\hat{g}} \hat{g}^{\mu_1 \nu_1} \hat{g}^{\mu_2 \nu_2} 
\hat{g}^{\mu_3 \nu_3} \hat{g}^{\mu_4 \nu_4}
f_{\mu_1 \mu_2 \mu_3 \mu_4} f_{\nu_1 \nu_2 \nu_3 \nu_4}. 
\label{36}
\end{eqnarray}
Then the normalized wave function is of form
\begin{eqnarray}
\hat{u}(z) &=& \sqrt{\frac{3 \omega}{2}} (\sin \omega z)^{\frac{3}{2}} \nn\\
&=& \sqrt{12 \omega} \frac{1}{(\e^{\omega r} + \e^{-\omega r})^{\frac{3}{2}}}
\nn\\
& \approx& \sqrt{12 \omega} \e^{- {\frac{3}{2}} \omega r},
\label{37}
\end{eqnarray}
Again, this normalized wave function indeed satisfies the Dirichlet 
boundary condition at $z = \frac{\pi}{\omega}$ and has an exponentially 
decreasing form $r \gg 1$. 

Thus far, we have studied p-form potentials ($p = 0, 1, 2, 3$) in five
dimensions.
This analysis can be extended to p-form potentials in higher dimensions in 
a perfectly analogous manner. Recently, we have constructed locally
localized gravity models in higher dimensions \cite{Oda6} where 
it has been pointed out that there is a nontrivial higher-dimensional
extension of Karch-Randall model \cite{Karch}. (There is also a trivial
extention of Karch-Randall model in higher dimensions, where the physical
features of the brane is the same as the Karch-Randall model so we will
not consider the trivial extension in this paper.) 
In the nontrivial model,
an $AdS_4$ brane is located at the origin $r=0$ in a general $D$-dimensional
space-time with negative cosmological constant. The line element has the
following form in the 'radial' $r$-coordinates and the 'conformal' 
$z$-coordinates
\begin{eqnarray}
ds_{(D)}^2 &=& g_{MN} dx^M dx^N  \nn\\
&=& \e^{-A(r)} \hat{g}_{\mu\nu} dx^\mu dx^\nu + dr^2  
+ R_0^2 \e^{-A(r)} d\Omega_{n-1}^2 \nn\\
&=& \e^{-A(z)} (\hat{g}_{\mu\nu} dx^\mu dx^\nu + dz^2
+ R_0^2 d\Omega_{n-1}^2), 
\label{38}
\end{eqnarray}
where $\e^{-A(r)}$ and $\e^{-A(z)}$ are given in Eq. (\ref{2}).
Here $\omega$, $R_0$, and the brane cosmological constant 
$\Lambda_{AdS}$ are expressed in terms of the bulk cosmological
constant $\Lambda$ and the vacuum expectation value $\eta$ of 
scalar fields as follows:
\begin{eqnarray}
\omega &=& \sqrt{\frac{-2 \Lambda}{(D-1)(D-2)}}, \nn\\
R_0^2 &=& \frac{1}{2 \Lambda} (D-2)(n-2-\kappa_D^2 \eta^2), \nn\\
\Lambda_{AdS} &=& \frac{2}{D-1} \Lambda,
\label{39}
\end{eqnarray}
where $n$ and $\kappa_D$ denote the number of extra dimensions and the
$D$-dimensional gravitational constant, respectively.

In particular, we have studied the localization property of p-form 
potentials in the case of $D=6, n=2$, that is, a string-like defect
model \cite{Oda3}. (Note that the analysis becomes more complicated 
as the number of $D$ gets large.) The results are very similar to
the case of an $AdS_4$ brane in $AdS_5$ as shown so far, so we will not
expose the detailed calculations any longer. 
The only difference we should mention
occurs in the case of a 2-form, which is self-dual in six dimensions.
Actually, provided that we start with the action (\ref{23}) in six
dimensions and assume a simple form of a solution 
$A_\mu(x^M) = a_\mu(x^\lambda) u(z) v(\theta), A_z(x^M) = A_\theta(x^M) = 0$ 
where $\theta$ is a sixth dimension 
denoting the angular variable in extra space, Eq. (\ref{27}) is changed
to $\partial_z^2  u = 0$ where we have considered the zero angular momentum,
that is, $\partial_\theta^2  v = 0$, for simplicity. Note that owing to the 
self-duality the exponential factor including $A(z)$ has disappeared.
The normalizable solution is given by $u(z) = u_0$. Then,
the normalized wave function becomes $\hat{u}(z) = \sqrt{\frac{2 \omega}
{\pi}}$, so it neither satisfies the Dirichlet boundary condition at
$z = \frac{\pi}{\omega}$ nor has an exponential damping factor.
However, there is a loophole in this argument. It is well known 
that there does not exist a covariant
action for the self-dual 2-form in six dimensions \cite{Marcus}, so
we cannot start with the action (\ref{23}) in the case at hand.
In other words, we cannot deal with the self-dual forms within the 
present approach.

In conclusion, we have studied the localization property of p-form
potentials on an $AdS_4$ brane in the locally localized gravity models.
Although we have shown that all the p-form potentials are trapped on
the brane, the corresponding wave functions spread rather widely in
a bulk space-time when the brane cosmological constant is small.
This fact seems to suggest that extra space is small enough not to
conflict with observations. It is of interest that the size of extra
dimensions is related to that of the cosmological constant. This
might shed new light on both the cosmological constant problem and
the compactification scenario.

\vs 1


\end{document}